# Electronic states of disordered grain boundaries in graphene prepared by chemical vapor deposition

Péter Nemes – Incze[1,2,*], Péter Vancsó[1,2], Zoltán Osváth[1,3], Géza I. Márk[1,2], Xiaozhan Jin[2,4], Yong-Sung Kim[2,6], Chanyong Hwang[2,4], Philippe Lambin[5], Claude Chapelier[3], László Péter Biró[1,2]

*1 Institute of Technical Physics and Materials Science, Centre for Natural Sciences, 1525 Budapest, PO Box 49, Hungary*

*2 Korean-Hungarian Joint Laboratory for Nanosciences (KHJLN), P.O. Box 49, 1525 Budapest, Hungary*

*3 SPSMS, UMR-E 9001, CEA-INAC/UJF-Grenoble 1, 17 Rue des Martyrs, 38054 GRENOBLE cedex 9, France*

*4 Center for Nano-metrology Division of Industrial Metrology Korea Research Institute of Standards and Science, Yuseong, Daejeon 305-340, South Korea*

*5 University of Namur (FUNDP), 61 Rue de Bruxelles, B-5000 Namur, Belgium*

*6 Center for Nanocharacterization, Division of Industrial Metrology, Korea Research Institute of Standards and Science, Yuseong, Daejeon 305-340, Republic of Korea*

Perturbations of the two dimensional carbon lattice of graphene, such as grain boundaries, have significant influence on the charge transport and mechanical properties of this material. Scanning tunneling microscopy measurements presented here show that localized states near the Dirac point dominate the local density of states of grain boundaries in graphene grown by chemical vapor deposition. Such low energy states are not reproduced by theoretical models which treat the grain boundaries as periodic dislocation-cores composed of pentagonal-heptagonal carbon rings. Using *ab initio* calculations, we have extended this model to include disorder, by introducing vacancies into a grain boundary consisting of periodic dislocation-cores. Within the framework of this model we were able to reproduce the measured density of states features. We present evidence that grain boundaries in graphene grown on copper incorporate a significant amount of disorder in the form of two-coordinated carbon atoms.

**I. Introduction**

Graphene prepared by chemical vapor deposition (CVD) on metal surfaces, especially copper [1], is becoming a material with strong potential for applications [2]. Perturbations of the two dimensional carbon lattice of graphene such as ripples [3] and grain boundaries (GB) [4] have significant influence on the properties of CVD graphene, from charge transport [5] to mechanical compliance [6]. Revealing the structure of GBs can go a long way to better understand the properties and behavior of CVD graphene and through this, enable the tailoring of GB properties.

---

\* Corresponding author
email: nemes.incze.peter@ttk.mta.hu

The physical properties of graphene GBs are usually computed by modeling them as periodic dislocation cores composed of non hexagonal units, mostly pentagons and heptagons [7–10]. Such structures have been observed previously on the surface of graphite (HOPG) using scanning tunneling microscopy (STM), clearly showing the pentagon – heptagon dislocation cores [11] and the periodic chain of local density of states (LDOS) perturbations caused by dislocation cores or other periodic defects [12–15]. These periodic dislocation cores are usually a low total energy configuration for a specific GB [7,9]. However, this does not necessarily mean that these structures will be present in realistic samples of CVD grown graphene, especially considering the very different growth conditions of HOPG [16] and polycrystalline graphene [17]. Indeed, various groups have started to calculate the properties of more realistic GBs, which incorporate carbon rings with four to nine members and structures containing twofold coordinated carbon atoms [6,18–20].

Due to the two-dimensional nature of graphene the atomic structure of its GBs is much more simple than in the case of bulk materials and lends itself much better to study by various microscopy methods. STM measurements together with theoretical calculations have been proven to be a reliable tool in determining the structure of GBs in graphite (HOPG) [11]. In the case of CVD grown graphene, transmission electron microscopy (TEM) is the tool of choice, mainly because of the availability of large area graphene and its transparency to electron beams. However, a certain degree of care needs to be taken while interpreting these results. Atomic resolution TEM images are usually measured at an electron energy of 80 keV [21–23] or 60 keV [24]. At this beam energy diffusion of defects and bond rotation at GBs is possible [22,25]. Various groups have reported pentagon – heptagon dislocation core structures for GBs [21,22,24], similar to the ones seen in graphite [11]. Others have observed aperiodic arrays of pentagons and heptagons [24] and more disordered grain junctions [23]. The differences may be due to sample variability, but also due to the modification of the sample under the electron beam. In contrast, STM does not perturb the structure of the GBs which is a significant advantage when interpreting the measurements.

In this paper we present a study of the atomic and electronic structure of graphene GBs, comparing

STM topography and spectroscopy data with *ab initio* density functional calculations. Our STM measurements show that localized states near the Dirac point, so called "midgap states" [26], dominate the LDOS of GBs. By extending the pentagon-heptagon dislocation core model [7,8] we have been able to reproduce the characteristic LDOS features of disordered GBs.

**II. Experimental and computational methods**

The graphene samples were prepared by the CVD method on polycrystalline copper foils in a similar process as our earlier results [17]. Briefly, the polycrystalline copper foils (Alfa Aesar) were annealed at 990°C for 30 minutes in a hydrogen atmosphere (2 sccm). The graphene growth was carried out by introducing $CH_4$ gas, at a flow rate of 35 sccm for 2.5 minutes. After graphene growth, the sample was allowed to cool to room temperature at a cooling rate of 50 K/min.

STM measurements were carried out at low temperature (4 K and 1.4 K) in a He cryostat, using a home built STM setup. The dI/dV spectra were recorded by the lock-in method, using a bias modulation of 2 mV. Typical parameters used for imaging were 0.3-1 nA tunneling current and 200-500 mV bias voltage.

We performed density functional theory (DFT) calculations within the framework of local density approximation (LDA) using the VASP simulation package [27,28] to investigate the electronic properties of the GBs. Projector augmented wave (PAW) pseudo-potentials [29,30] were used and the kinetic energy cut-off for the plane wave expansion was 400 eV. We used a rectangular super-cell with two parallel GBs to achieve periodical boundary conditions (see also Ref [7]). The dimension of the super-cell was 39.2 Å across the GBs and 6.5 Å along the GBs. We verified that a larger super-cell of 39.2x13 Å produces only negligible changes in the calculated atomic and electronic structure of the GBs. In all geometries atomic positions were relaxed using the conjugate–gradient method until the forces of the atoms were reduced to 0.03 eV/Å, and the Brillouin-zone was sampled using a 2x12x1 Γ-centered k-space grid. STM images were simulated with the simple Tersoff-Hamann approximation [31] using the calculated local density of states.

**III. Results and discussion**

    *III.1. STM measurements*

After graphene growth, the stepped surface of the copper is covered by graphene, the measurement of the GBs was done on the copper terraces (see Fig. 1). GBs can be identified as lines of protrusions on the graphene surface, while subsequent atomic resolution imaging was used to determine the mismatch angles (see insets in Fig. 2d). In Figure 2a-c we show topographic STM images of various GBs. The protrusions at the boundary can be attributed to two effects. They can be created by out of plane buckling of the graphene at the GB, due to the relaxation of mechanical stress [10]. Furthermore, electronic states localized at the GB can cause an increase in the tunneling current, leading to an increase in the measured height of the boundary. This height can be 0.6 nm or larger in certain places (see inset in fig. 2c). This is at least twice as large as the expected out of plane buckling in free space of GBs due to the relaxation of mechanical stress [10]. Therefore, some of the observed height can be attributed to localized electronic states of the specific atomic configuration at the GB [7]. What is striking about these measurements is that in the case of a GB with a periodic array of dislocation cores, we would expect a much more uniform and possibly periodic distribution of LDOS induced corrugation, as in the case of measurements on HOPG [11–14]. Instead we observe that the corrugation of the GBs is irregular and disordered, with only traces of periodicity (see Fig. 2a). Examining the GBs in atomic resolution we can find further evidence that points to a more disordered structure. Figure 2e shows a constant current STM image of the graphene lattice near a GB (Fig 2d). Superimposed on the atomic corrugation is a higher wavelength pattern which has a periodicity of $\sqrt{3}$ times that of graphene. The presence of this so called $\sqrt{3} \times \sqrt{3}R30^o$ type superstructure is also apparent in the Fourier transform of the image, where the corresponding peaks are located at smaller wave-numbers and are rotated by 30° relative to the peaks of the graphene reciprocal lattice (Fig. 2f). This superstructure is caused by the scattering and interference of electronic states close to the Fermi level. It is a typical hallmark of inter-valley scattering in graphitic systems [32]. A strong source of this type of scattering are sharp lattice defects, for example under-coordinated atoms [33], which could also contribute to the appearance of weak localization in magneto-transport measurements of individual GBs [4]. Further evidence to the nature of these lattice defects can be obtained by investigating the LDOS of the GBs.

In Figure 3a we present dI/dV spectra at points along the GB shown in Figure 3b. A frequent feature of the spectra is a strong localized peak near 0 V bias. This peak is most apparent if we measure the dI/dV spectra on top of the protrusions seen in the topographic images. It is a common feature of all the GBs we have measured. In order to show that such LDOS peaks are indeed localized in the vicinity of the GB we plot a line of dI/dV spectra across a GB using a color map to show the magnitude of the tunneling conductance (see Fig. 4). In the region of the dI/dV plot measured on the graphene surface far from the GB, we can observe a clear minimum of the spectra at the slightly positive bias voltage of 0.15 V. In the case of the clean graphene surface, we interpret this as the minimum of the graphene density of states at the Dirac point [34,35]. The shift of the LDOS features and LDOS minimum towards a positive bias voltage means that the graphene is p-doped. This doping was observed throughout our measurements independent of the measurement location on the sample. Such doping can arise due to the difference in work function between the graphene and copper, by a weakly coupled graphene – copper system [36]. As the tip is approaching the GB we can observe the appearance of the localized state at 0.1 – 0.15 V, close to the Dirac point position (white ellipse in Fig. 4).

To better demonstrate that the localized peaks close to the Dirac point are a frequent presence in the LDOS of GBs, we show a histogram of the LDOS peak positions from 60 individual dI/dV curves measured on top of various GBs (see Fig. 3d). The peaks around 0.4 eV can be produced by certain pentagon – heptagon dislocation cores [7], but the overwhelming majority of the peaks is present in a 0.1 eV interval close to the charge neutrality point. The observation that these localized states are present in all GBs suggests that the atomic structure of boundaries with different mismatch angles has common aspects. This behavior is particularly interesting, since for GBs with different mismatch angles the low energy dislocation core structure is different [7]. Thus, we would expect differing peaks in the LDOS for boundaries with different mismatch angles.

In graphene a strong localized state appears close to the Dirac point if the lattice has defects with more atoms on one sublattice than the other, for example a zigzag edge [37] or edges and cracks with zigzag edged components [38,39]. Furthermore, vacancies and divacancies have a very similar signature [35,40–42]. Using tight binding calculations, Carpenter et al. have calculated the DOS of graphene with high concentrations of vacancies, an almost amorphous system [43]. At all vacancy concentrations studied they have observed the strong peak at the charge neutrality point. Exotic non hexagonal ring structures can have states close to the Dirac point as well [44,45], but they are not likely to be the major factor in producing the peaks observed here. This is because the formation of some of these structures would have to be highly preferential for it to be a defining feature of the GB LDOS. Furthermore, atomic resolution STM images of line defects composed of purely non-hexagonal units, without strong lattice defects, do not show signs of inter-valley scattering [11,44].

It is reasonable to expect disorder at the GBs of CVD graphene and structural differences from the low energy dislocation core configurations which have been found in HOPG [11]. The reason behind this difference can be found if we consider the synthesis conditions for CVD graphene and HOPG. In the case of HOPG, the preparation involves temperatures in excess of 3000°C and annealing times of 30 to 60 minutes following the growth phase [16]. Such high temperatures are known to reduce defects in graphitic systems [16], allowing the GBs to reach their low energy state. In the case of CVD graphene both the annealing temperature and time are much less. The temperature is in the range of 1000°C and the growth time is measured in seconds [17] to tens of minutes [1,2]. In the latter case the lower temperature may not be sufficient to adequately fuse the growing graphene domains, which can leave behind vacancies and other structures containing two-coordinated carbon atoms.

### *III.2. DFT calculations of disordered grain boundaries*

In order to study realistic GBs, we have extended the pentagon-heptagon dislocation core model [7] to include disorder. This was achieved by introducing vacancies in a GB and allowing the structure to relax, using DFT based molecular dynamics. Due to the large computational cost of the DFT relaxation calculations, we could not calculate all the various GB geometries encountered during measurements. Instead, we have chosen a sampling method. As the starting geometry we investigate a model system composed of 5-7 carbon rings arranged periodically (Fig. 5a). This system has been extensively studied in the literature [7,8,11] and has a similarly high mismatch angle (21.8°) as the GBs, presented in this paper. Within the DFT unit cell, we have systematically removed carbon atoms from all possible lattice positions at the GB, marked by numbered carbon atoms in Fig. 5a. The atomic structure of the six GBs sampled in this way, was obtained after DFT relaxation (see Fig. 5c). All these structures are found to contain non hexagonal carbon rings and two coordinated sites. It is worth mentioning that the carbon atom configurations of these boundaries are remarkably similar to the ones resulting from computationally less expensive methods, such as the one used by Malola et al [19]. However, in our case VASP calculations allow us to study the electronic states of these structures in much greater detail. In Fig. 5b we have plotted the DOS of the six GB configurations, as well as that of the unperturbed 5-7 system. The DOS of the 5-7 GB matches the results of Yazyev et al. [7], showing peaks in the vicinity of ±0.4 eV. The rest of the DOS curves, corresponding to the six disordered GBs, show peaks at various energies. Common features are strong peaks in an energy interval of 0.2 eV around the Fermi level. However, the peak positions of the six disordered structures have the low energy DOS peaks at slightly different energies. From the histogram of peak positions in Fig. 3d it is clear that a similar variability is also present in the measured dI/dV spectra. Thus, the variability of the positions of the low energy LDOS peaks can be attributed to slightly different local atomic configurations of the disordered grain boundary. A better approximation of the measured LDOS peaks can be obtained if we calculate a statistical average based on the DOS of the six different GB configurations. To achieve this we need to calculate the formation energy of the GBs. The formation energy per unit length ($\varepsilon_i$) of the $i^{th}$ GB, relative to the pristine GB is defined as follows: $\varepsilon_i = (E_i - E_{5-7} + 2E_{C\,atom})/2L$, where $E_i$ and $E_{5-7}$ is the energy of the $i^{th}$ disordered GB and the pristine 5-7 GB respectively, over the whole DFT super-cell and $E_{C\,atom}$ is the energy per carbon atom for the pristine graphene. Because the super-cell contains two GBs due to periodicity requirements, we have to divide by twice the unit cell length along the GBs (L). Additionally, we account for the fact that the disordered GBs contain two less carbon atoms because of the added vacancies. Thus, the formation energies of the disordered GBs from 1 to 6 are as follows: 10.41, 9.82, 11.13, 10.02, 10.77 and 11.43 eV/nm, with GB no. 2 having the lowest formation energy. The formation energy of the pristine 5-7 GB per unit length is found to be 4.03 eV/nm, in good agreement with results from the literature [46,47].

The statistical average of all the calculated GB configurations can be seen in Fig. 3c, plotted on the same energy scale as the measured data to make it easier to compare the two. Comparing this result with the measured LDOS peak positions we can conclude that by introducing disorder in the 5-7 periodic GB we could reproduce the low energy LDOS peaks observed in measurements. From the DFT results we can also conclude that the low energy states are localized on and around the two coordinated atomic sites. This is apparent from the contour plot of the density of low energy states for a disordered GB (Fig. 6a). A closer inspection of the calculated DOS (Fig. 5b) reveals that in most cases both $sp^2$ and $p_z$ type states contribute to the formation of the peaks close to the Dirac point. Tight binding calculations, which take into account only π orbitals, have already predicted the presence of low energy peaks in the DOS of defects containing two-coordinated carbon atoms [20,40]. Our results show that $sp^2$ dangling bonds can give a dominant contribution to the low energy states of these defects.

During STM measurements the positions of the carbon atoms at the GB could not be resolved, so that we have no direct atomic resolution images of two-coordinated atoms in the GB structure. The explanation of the lack of atomic resolution on the disordered GB can be explained with the help of the calculated DOS. It is clear from Fig. 6a that the low energy states are not localized exactly on the atomic positions. Furthermore, the intensity of these DOS peaks is by far the largest in both the calculated and measured DOS. This means that their contribution to the STM current will be large, which together with the finite curvature of the STM tip leads to a smearing effect of the LDOS features. This smearing is evident from the simulated STM image in Fig. 6b, where the atomic sites at the boundary can not be distinguished. A high defect concentration at the GB implies that the atomic corrugation would be smeared out in STM images, leaving them as lines of protrusions. It should be pointed out that the lateral size of the GB may be larger than the geometries we have considered in our model. Depending on the concentration of lattice defects, a GB could contain more than one under-coordinated carbon atom along its width, thus increasing its lateral dimensions. Using classical molecular dynamics Kotakoski et al. have explored the mechanical properties of such highly disordered GBs [6] and have successfully reproduced the behavior of CVD graphene subjected to mechanical stress. Indeed, it is likely that depending on the graphene growth conditions, CVD graphene can contain a varying density of defects along the GBs. However, since the midgap states are localized on the two-coordinated sites and show only slight variability in energy depending on the atomic neighborhood, we would not expect significant differences in the LDOS if a wider GB is considered.

As a final remark we would like to point out that even though we have considered only a limited number of GB geometries, generally applicable conclusions can be drawn from these. In our calculations, disordered GBs were generated by introducing vacancies into a perfect GB composed of pairs of 5-7 membered carbon rings. The atomic structure of each resulting GB contained two-coordinated atoms. Similar structures were found using a different approach, that of growing disordered GBs atom by atom, using molecular dynamics [5,6,19,20]. Therefore, it is safe to assume that two-coordinated atoms are ubiquitous in disordered GBs. We have shown that such a severe disruption of the aromatic carbon ring structure gives rise to peaks near the Dirac point energy, which are localized at the sites of two-coordinated carbon atoms. These midgap states have similar energy as the localized states of point defects in graphene [40]. However, disordered GBs can not be considered as simple lines of vacancies. This is because there is a slight variability in the energy of the midgap states of ~0.1 eV, depending on which GB atomic structure is considered. Similar variability is also observed in the energy of the dI/dV peaks in the measurements. Furthermore, in addition to the midgap states, disordered GBs have DOS peaks at higher energy.

## IV. Conclusions:

Using STM we have been able to directly probe the mismatch angles and the electronic properties of GBs. The ubiquitous signature of inter-valley scattering and the presence of states near the Dirac point lead us to conclude that disorder in the form of two-coordinated carbon atoms plays a significant role in defining the electronic properties of GBs. We have extended the periodic dislocation-core model to include disorder of this kind and have reproduced the main DOS features observed by STM. We expect that inter-valley scattering at these defect sites is a leading influence on charge transport phenomena. For example it can contribute to weak localization in the case of inter-grain transport [4] and may destroy the high transparency [8] of certain dislocation core configurations to charge carriers. Furthermore, the predicted exotic properties [8,44] of GBs may only be accessible to experimental investigation if disorder can be eliminated to a sufficient degree. This could be achieved by better control of the growth parameters [48]. At the same time, the increased density of under-coordinated atoms at the boundary could prove to be a better source of carbon magnetism [18] than the GBs of HOPG [13].

**Acknowledgments**

This work has been conducted within the framework of the Korean-Hungarian Joint Laboratory for Nanosciences (KHJLN), the Converging Research Center Program through the Ministry of Education, Science and Technology (2010K000980), OTKA grant PD 84244 and K 101599. Furthermore, the authors wish to acknowledge the support of the French-Hungarian bilateral program PHC-Balaton/TÉT_10-1-2011-0752 and the French ANR-BLANC-SIMI10-LS-100617-12-01 grant. GIM is grateful for the support from the EU Marie Curie International Research Staff Exchange Scheme Fellowship within the 7th European Community Framework Programme (MC-IRSES proposal 318617 FAEMCAR project)

**Figures:**

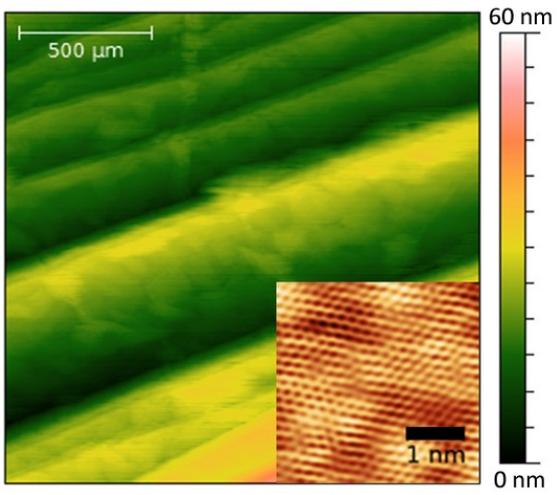

*Figure 1. STM measurement of the stepped surface of the copper substrate. Inset shows the hexagonal atomic lattice of graphene on top of a copper terrace.*

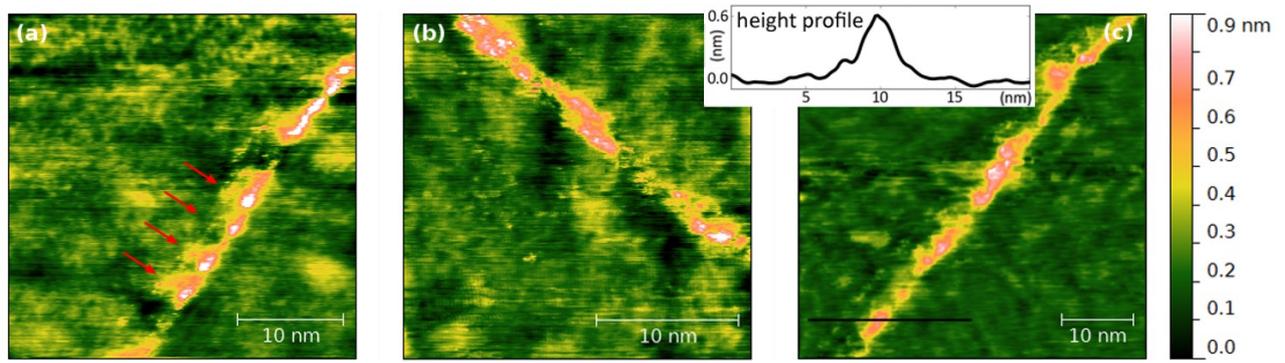
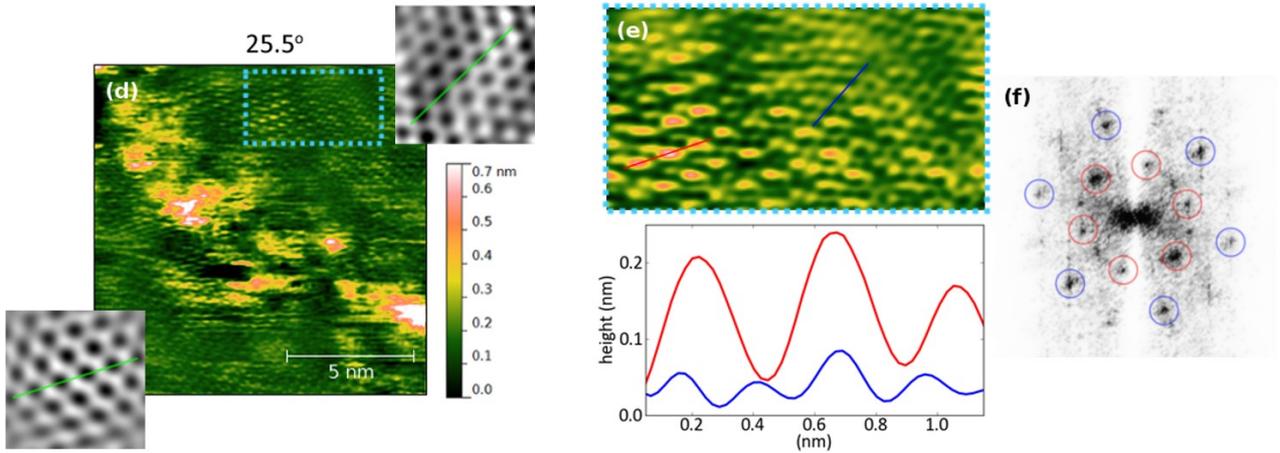

*Figure 2. **(a, b)** STM topography of two different GBs. **(a, c)** Two different segments of the same boundary. The GBs in CVD graphene usually show a disordered array of protrusions. **(a)** Occasionally periodic structures can be observed. **(d)** STM topography image of a disordered GB, with a mismatch angle of 25.5°, showing $\sqrt{3} \times \sqrt{3} R30^o$ type superstructures. Insets: atomic resolution STM image on the two graphene domains, green lines show the zigzag direction. **(e)** Zoomed in image showing the superstructure and height profiles corresponding to the red (superstructure) and blue (atomic structure) lines in (e). **(f)** Fourier transform of the atomic resolution image in (d). Colored circles are used to highlight the peaks corresponding to the atomic lattice (blue) and the superstructure (red) of one of the graphene grains.*

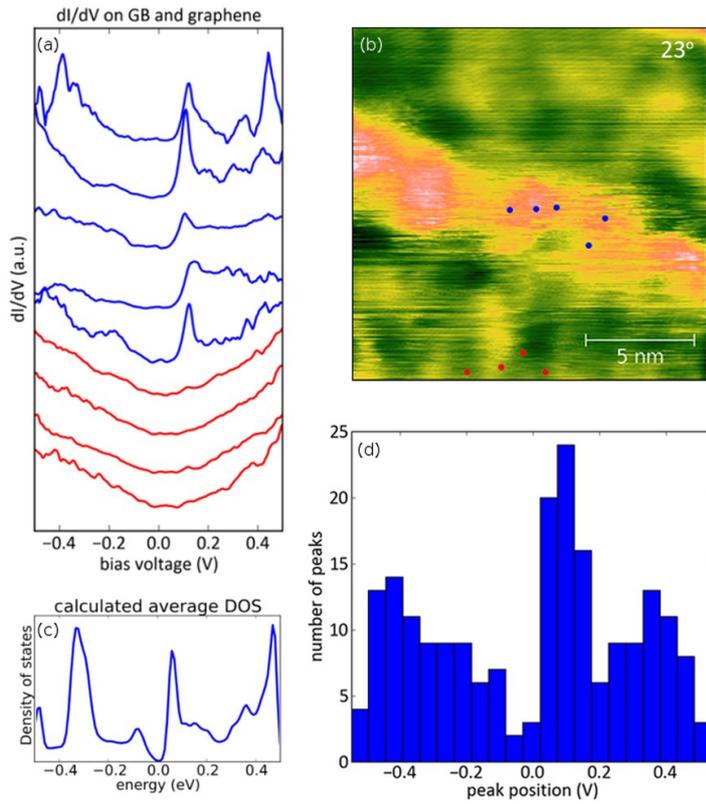

*Figure 3. **(a)** dI/dV plots measured at selected points shown on the topography image. **(b)** Blue curves were measured on the GB and red curves on the graphene, the positions of the measurement are shown by similarly colored dots. **(c)** statistical average of the density of states of the six disordered GBs, calculated by DFT. **(d)** Histogram of the positions of LDOS peaks in 60 dI/dV spectra, measured over different positions on three distinct GBs (misorientation angles: 23º, 25.5º, 27º).*

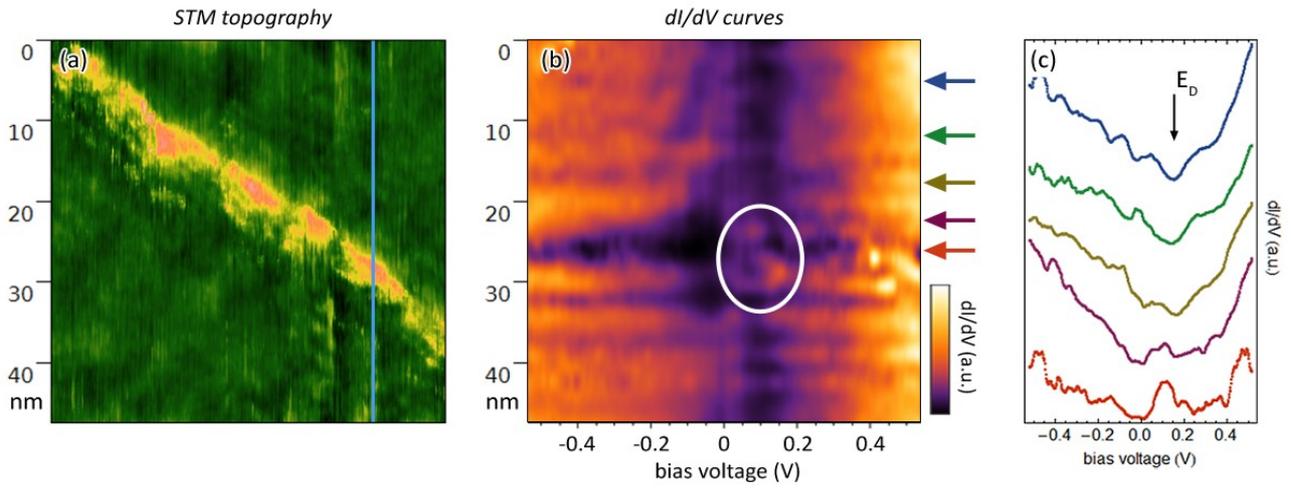

*Figure 4. **(a)** STM topography image of a GB. A line of dI/dV spectra was measured across the GB in the position shown by the blue line. **(b)** False color plot of the line of dI/dV spectra across the GB. White ellipse shows the localized states close to the Dirac point. **(c)** Selected dI/dV spectra from positions shown by appropriately colored arrows in (b).*

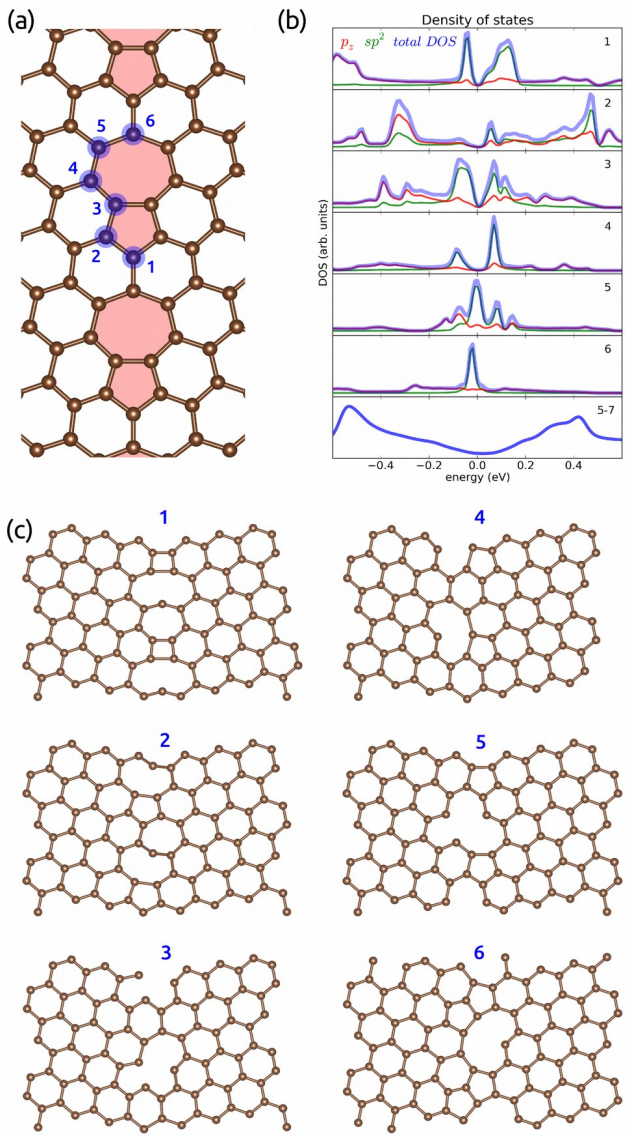

*Figure 5. **(a)** GB constructed from 5-7 membered carbon rings, with a mismatch angle of 21.8º. Removed carbon atoms are numbered. **(b)** DOS of the unperturbed 5-7 boundary and of the six disordered structures produced by inducing vacancies in the 5-7 GB. Fermi level is at 0 eV. **(c)** Atomic positions of the six disordered GBs.*

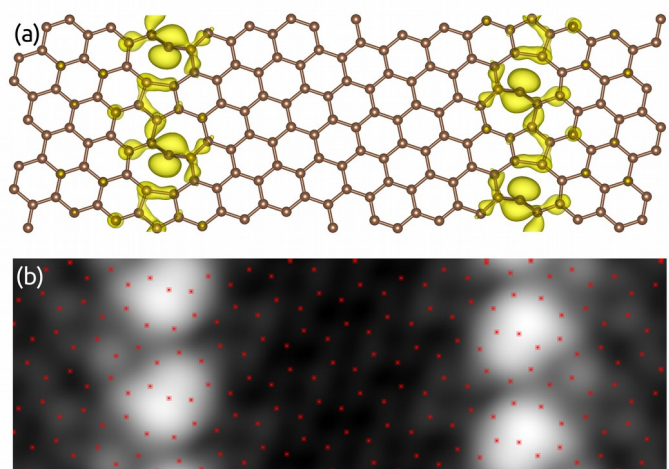

*Figure 6. **(a)** Calculated local density of states in the 0 to 0.2 eV energy interval, for the disordered GB nr. 2. **(b)** Simulated STM image based on the DFT DOS data for the same grain boundary. Red dots show the carbon atom positions.*